# Spiral Galaxies as Enantiomers: Chirality, an Underlying Feature in Chemistry and Astrophysics

SALVATORE CAPOZZIELLO[a] AND ALESSANDRA LATTANZI[b]

[a] *Dipartimento di Scienze Fisiche e INFN sez. di Napoli, Università di Napoli "Federico II", Complesso Universitario di Monte S. Angelo, Via Cinthia I-80126, Napoli, Italy*
[b]*Dipartimento di Chimica, Università di Salerno, Via S. Allende, 84081, Baronissi, Salerno, Italy*
E-mail: capozziello@na.infn.it; lattanzi@unisa.it

*ABSTRACT* Spiral galaxies are axi-symmetric objects showing 2D-chirality when projected onto a plane. Features in common with tetrahedral molecules are pointed out, in particular the existence of a preferred chiral modality for genetic galaxies as in aminoacids and sugars. Environmental effects can influence the intrinsic chirality of originally isolated stellar systems so that a progressive loss of chirality is recognised in the Hubble morphological sequence of galaxies.

## INTRODUCTION

As a general definition, enantiomers are objects or physical systems whose mirror images are not superimposable as hands and shoes. In chemistry, they are molecules with the same formula, the same atomic connectivity and the same distances between corresponding atoms, but with non-superimposable mirror images. This feature is easily recognized in a wide range of natural systems and it is generally reported under the standard of chiral symmetry. While it is well-known that elementary particles (e. g. L-neutrinos) or organic molecules (e. g. L-aminoacids and D-sugars) have a preferential chiral characterization, it has to be assessed if extremely large macroscopic systems can share this feature. Spiral galaxies are among these systems and it is important to stress that chirality could be the *imprint* of some primordial microscopic evolutionary process, which led to the today observed cosmic large scale structure. In fact, a commonly accepted paradigm in cosmology asserts that initial quantum fluctuations have been hugely enhanced during the inflationary epoch, giving rise to the observed galaxies and clusters of galaxies, which are the largest structures in the Universe.[1] The order of magnitude of such an expansion ranges from $10^{50}$ to $10^{60}$ so that, in principle, every microscopic process could have been enlarged to huge astrophysical scales.[1] On the other hand, chirality of elementary particles plays a fundamental role in physics being related to symmetry breaking which led to the violation of some fundamental conservation laws as *CP* invariance. It is therefore legitimate to ask for remnants of these processes not only in microscopic phenomena, but also in macroscopic ones, due to the cosmological expansion, with the perspective of a unitary view of physics.[2]

In this contribution, we develop some considerations by which spiral galaxies can be framed under the standard of molecular chirality. Some recent studies have shown that chiral tetrahedral molecules are algebraically described by orthogonal groups[3] and the approach gives evidence of common features with those of elementary particles. In particular, molecules and particles can be represented by unitary quaternions with analogous dynamics.[4] Spiral galaxies are endowed with these features and a defined chiral modality could be a genetic property which galaxies are going to loose due to environmental effects.

The article is organized as follows: in Sec. 2, orthogonal groups are discussed, showing as they are naturally suitable to describe chiral objects. In particular, tetrahedral molecules and particles can be described by the *O*(4) group. Sec. 3 is devoted to the spiral structure of galaxies. From an algebraic point of view, their chiral features as trailing and leading arms are recovered by the *O*(2) group. Environmental effects and evolution toward achiral systems in the Hubble sequence are discussed in Sec. 4. An outlook of the discussed concept is given in Sec. 5.

## CHIRALITY BY ORTHOGONAL GROUPS

In any way it is conceived, chirality is always related to the fact that structurally identical objects have non-superimposable mirror images. According to Ruch,[5] we have a genuine form of chirality, the so-called *handed* chirality (shoe-like) and the so-called *nonhanded chirality* (potato-like). In the case of tetrahedral molecules, enantiomers are handed chiral objects, while octahedral complexes, having six different ligands, can be an example of nonhanded chiral objects.[6] In general, we deal with not superimposable mirror images as chiral objects (enantiomers), neither superimposable nor mirror images chiral objects (diastereoisomers), and achiral objects (isomers which are completely superimposable by rotation). In any case, the chiral nature of objects can be established taking into account the group of transformations acting on the configuration space, i.e. the *ensemble* of all the possible configurations of the system itself. In this sense, chirality is a symmetry which emerges in abstract space (e.g. the spin-helicity configuration space of elementary particles) and it depends on the true structure of the object, rather than on the physical space where it is embedded. With this concept in mind, if a system is constituted by N elements, the group of possible transformations is $G(N)$, where $G$ is defined in the abstract space of dimension N.[3] A general definition of chirality is then related to the number of possible rotations and inversions which a system with N constituents can undergo. It is then straightforward to consider the $O(N)$ groups which are the groups of rotations and inversions of a given system. The case of tetrahedral molecules is enlightening (Figure 1).

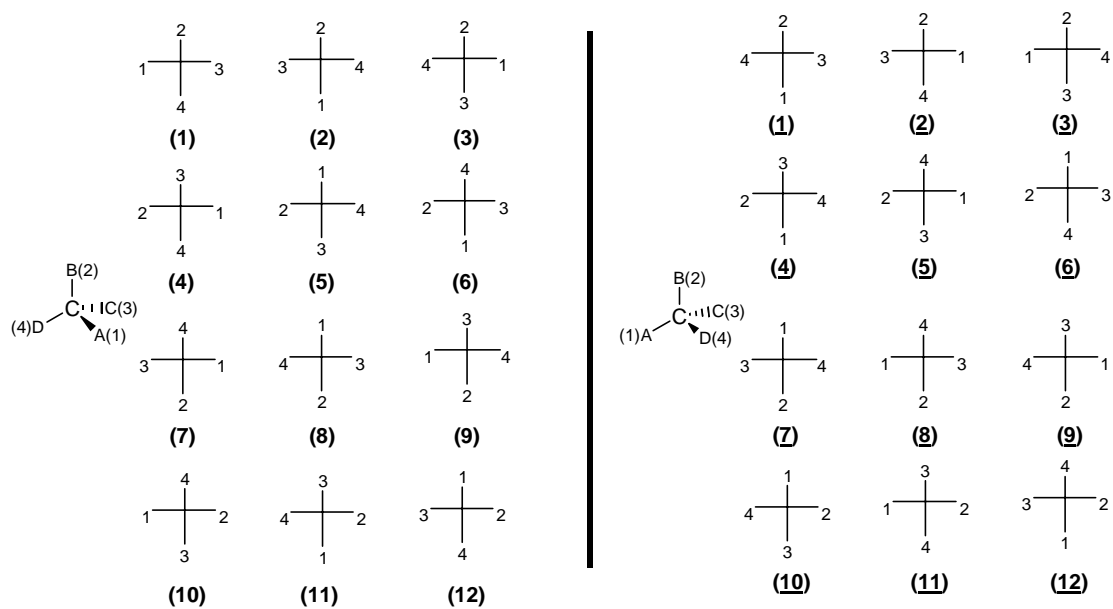

**Fig 1.** 24 Fischer projections for the enantiomers of a tetrahedral molecule.

A tetrahedron, constituted by 4 different ligands, has 4!=24 possible configurations which are nothing else but the Fischer projections of the two enantiomers of the molecule.[7]

A tetrahedral molecule can be assigned by a column vector $\Psi$,

$$\Psi = \begin{pmatrix} \Psi_1 \\ \Psi_2 \\ \Psi_3 \\ \Psi_4 \end{pmatrix} \quad (1)$$

where $\Psi_1, \ldots \Psi_4$ represent the bonds and the corresponding initial Fischer projection is **(1)** (Figure 1). The matrix representation of the projection **(1)** is assumed as fundamental, i.e. the identity matrix

$$\chi_1 = \begin{pmatrix} 1 & 0 & 0 & 0 \\ 0 & 1 & 0 & 0 \\ 0 & 0 & 1 & 0 \\ 0 & 0 & 0 & 1 \end{pmatrix} \quad (2)$$

and the action on the column vector $\Psi$ is nothing else but the identity. When acting with the transformations of the $O(4)$ group on the molecule, there can be rotations (of the same enantiomer) and inversions of a couple of bonds (the other enantiomer is obtained). Configuration **(2)** (Figure 1) can be achieved from the starting **(1)** as soon as an operator $\chi_2$ is defined acting as

$$\chi_2 \begin{pmatrix} \Psi_1 \\ \Psi_2 \\ \Psi_3 \\ \Psi_4 \end{pmatrix} = \begin{pmatrix} \Psi_3 \\ \Psi_2 \\ \Psi_4 \\ \Psi_1 \end{pmatrix} \quad (3)$$

which corresponds to the matrix

$$\chi_2 = \begin{pmatrix} 0 & 0 & 1 & 0 \\ 0 & 1 & 0 & 0 \\ 0 & 0 & 0 & 1 \\ 1 & 0 & 0 & 0 \end{pmatrix} \quad (4)$$

$\chi_2$ interchanges two of the bonds and it is equivalent to a rotation. The configuration (**1**) of the other enantiomer can be obtained starting from **(1)** as soon as an operator $\bar{\chi}_1$ is defined acting as

$$\bar{\chi}_1 \begin{pmatrix} \Psi_1 \\ \Psi_2 \\ \Psi_3 \\ \Psi_4 \end{pmatrix} = \begin{pmatrix} \Psi_4 \\ \Psi_2 \\ \Psi_3 \\ \Psi_1 \end{pmatrix} \quad (5)$$

which corresponds to the matrix

$$\bar{\chi}_1 = \begin{pmatrix} 0 & 0 & 0 & 1 \\ 0 & 1 & 0 & 0 \\ 0 & 0 & 1 & 0 \\ 1 & 0 & 0 & 0 \end{pmatrix}. \tag{6}$$

$\bar{\chi}_1$ generates the inversion between the bonds $\Psi_1$ and $\Psi_4$. By the same procedure, all the configurations in Figure 1 can be obtained.[3] The same enantiomer is conserved when acting on the molecule only with the elements of the sub-group $SO(4)$, the 4D-special orthogonal group.[3] The orthogonality conditions are $\frac{1}{2}4(4+1) = 10$, while the number of independent generators (parameters) of the group is $\frac{1}{2}4(4-1) = 6$. With this approach in mind, a chiral transformation is related to the possibility that two bonds of the molecule undergo an inversion, which, in the set of possible configurations, means a transition from the (+)-enantiomer to the (-)-enantiomer and vice-versa. From a quantum mechanical viewpoint, this can be a quantum tunneling process so that $O(4)$ represents a *quantum chiral algebra*.[3,4]

It is interesting to point out that the isomorphism

$$SU(2) \times SU(2)/\pm 1 \approx SO(4) \tag{7}$$

holds for groups $SU(2)$ and $SO(4)$ so that rotations and inversions of ligands can be given in terms of Pauli matrices or unitary quaternions.[4] Features like these indicate that quantum states of chiral tetrahedral molecules and fundamental particles show analogies which have to be seriously taken into account as previously suggested by Heisenberg.[2]

These arguments can be easily generalized to generic systems with an even number N of constituents (e.g. ligands), with N! configurations. The algebra will be $O(N)$, the number of independent generators will be $\frac{N(N-1)}{2}$. Clearly the word *enantiomer* has the general meaning of a physical system, present in two modalities, with non-superimposable mirror images.

## SPIRAL STRUCTURE: TRAILING AND LEADING ARMS

The above discussion can be specified for spiral galaxies which have their arms winding in a trailing or in a leading sense (Figure 2). *Trailing* means that the tips of the arms point in the opposite direction from rotation, while in the case of *leading* arms the tips point in the same direction of rotation. These features are not always immediately evident in the galaxy structure: they appear very clearly in the so-called *grand-design spirals*, which usually have two symmetric and well-defined arms as in the case of the galaxy M51. However, not all spirals are grand-design with two distinct arms; they can be *flocculent* and in this case the spiral structure is not simple to be recognised.

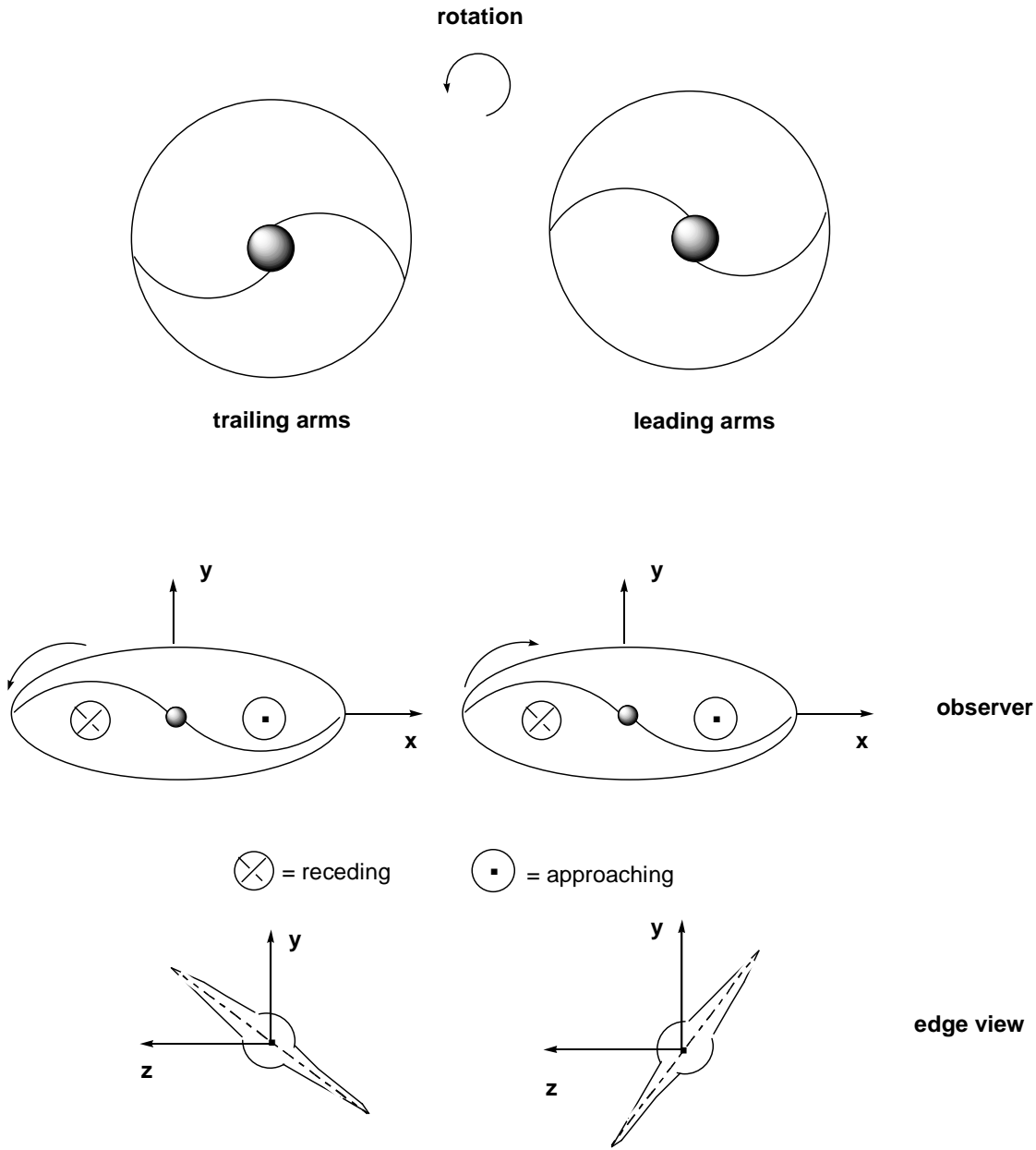

**Fig. 2.** Trailing and leading modes of spiral galaxies.

The distinction between trailing and leading spiral arms requires the determination of the orientation of the plane of the galaxy relative to our line of sigth (Figure 2), so that radial velocity measurements can be unambiguously interpreted in terms of the direction of galaxy rotation. Considering the recession velocity, that is given by the Hubble relation (which holds for every galaxy thanks to the cosmological expansion) $v = H_0 d$, where $H_0$ is the Hubble constant (today esteemed ~71 Km·sec$^{-1}$·Mpc$^{-1}$)[8] and *d* the distance of the galaxy from the observer, a definite chirality is assigned. Moving along the arms of the galaxy toward the center and taking into account the direction of recession velocity, the helicity is assigned so that trailing galaxies are left-helical, while leading galaxies are right-helical. However, although the trailing and leading arm objects at the top of Figure 2 are distinct and non-superimposable with respect to reflection within the plane, the sense of rotation has to be taken into account. In fact, the rotation will reverse under reflection within the plane. The enantiomers in two dimensions are, in fact, the trailing arms galaxy rotating in one sense (top of Figure 2) and the leading arms galaxy rotating in the opposite sense, which also makes it a trailing arms galaxy! Due to this problem, it is essential an absolute observational determination of the rotation of a galaxy with respect to a given observer (i.e. with respect to us). In principle, rotation of a galaxy is a "kinematic" quantity which cannot be reversed, thanks to the conservation of angular momentum and to the fact that galaxies, as we said, participate to the recession motion of cosmos with respect to us, unless the system undergoes some strong dynamical interaction, as it will be discussed below. This fact allows an absolute determination of trailing and leading modes. In order to be clear, it is worth entering into some details of observations.

Considering the two galaxies in Figure 2, in both cases, the (x,y)-plane is the celestial sphere and the z-axis is assumed to point toward the Sun. The galaxy on the left is inclined so that the side nearest the Sun is in the half plane y > 0, while the galaxy on the right has its near side in the half plane y < 0. In Figure 2, the spiral pattern and the rotation direction are indicated for both galaxies; the spiral galaxy on the left is leading and that on the right is trailing. Despite this difference, the appearance of both galaxies as seen from the Sun is the same; as shown in Figure 2, in both systems the spiral pattern appears to curve in an anti-clock-wise direction as one moves out from the center, and the side with x > 0 has radial velocity toward the Sun. Thus, radial velocity measurements and photographs of the spiral pattern cannot, by themselves, distinguish leading and trailing spirals. To determine whether a particular galaxy leads or trails, it must be determined which side of it is closer to us. One way to do this, is to count the numbers of objects such as standard candles as novae and globular clusters that are seen on either side of the apparent major axis. If the objects are heavily obscured by dust in the central plane, then fewer objects will be seen on the near side of the galaxy and this fact constitutes a problem. Apart from using standard candles, the near side can also be determined by studying the pattern of absorption in the dust lanes.[11] Summing up all these observational methods, unambiguous answers can be achieved.

In almost all cases, where this determination can be made, *it does appear that spiral arms are trailing*,[9] while the rare leading spirals are interpreted as the result of tidal encounters with retrograde-moving objects (the case of Andromeda galaxy M31 and its companion M32 is a remarkable example). In the next section, environmental effects will be discussed.

Turning to the orthogonal group description of chirality, the spiral arms can be considered as the ligands and the configuration space is 2D. The 2D-image (the galaxy proiected onto its lying plane) cannot be superimposed with its mirror image without removing the image from the plane. The two images are the leading and trailing structures sketched in Figure 2. The two modalities are parametrized by a rotation angle, then chirality is algebraically described by the $O(2)$ group, where the number of generators is $\frac{2(2-1)}{2}=1$, corresponding to the rotation angle. To demonstrate these statements, we can write any rotation into the plane as

$$\begin{pmatrix} y' \\ x' \end{pmatrix} = \begin{pmatrix} \cos\vartheta & \sin\vartheta \\ -\sin\vartheta & \cos\vartheta \end{pmatrix} \begin{pmatrix} x \\ y \end{pmatrix} \quad (8)$$

or, abbreviating,

$$x^{i'} = O^{ij}(\vartheta) x^j \quad (9)$$

with $x^1 = x$ and $x^2 = y$ where $O^{ij}(\vartheta)$ is the rotation matrix. Being the group orthogonal, we have $O^{-1}(\vartheta) = O(-\vartheta)$ and then

$$O^{-1}(\vartheta)O(-\vartheta) = \mathbf{I} = \begin{pmatrix} 1 & 0 \\ 0 & 1 \end{pmatrix} \quad (10)$$

which means that the inverse of an orthogonal matrix is its transpose. Any orthogonal matrix can be written as the exponential of a single antisymmetric matrix $\tau$:

$$O(\vartheta) = e^{\vartheta \cdot \tau} \equiv \sum_{n=0}^{\infty} \frac{1}{n!}(\vartheta\tau)^n \quad (11)$$

where

$$\tau = \begin{pmatrix} 0 & 1 \\ -1 & 0 \end{pmatrix} \quad (12)$$

so then we have that the transpose of $e^{\vartheta \cdot \tau}$ is $e^{-\vartheta \cdot \tau}$:

$$O^T = (e^{\vartheta \cdot \tau})^T = e^{-\vartheta \cdot \tau} = O^{-1} \quad (13)$$

Another way to prove this identity is to power expand the right-hand side and sum up the series. We obtain

$$e^{\vartheta \cdot \tau} = \cos\vartheta \cdot \mathbf{I} + \tau \sin\vartheta = \begin{pmatrix} \cos\vartheta & \sin\vartheta \\ -\sin\vartheta & \cos\vartheta \end{pmatrix} \quad (14)$$

which demonstrates that all elements of $O(2)$ are parametrized by one angle $\vartheta$, so we can say that $O(2)$ is a one-parameter group.

Let us now take the determinant of both sides of the defining Eq. [13]:

$$\det(OO^T) = \det O \det O^T = (\det O)^2 = 1 \tag{15}$$

which means that the determinant of $O$ is equal to $\pm 1$. If we take $\det(O) = 1$, the resulting subgroup is *SO*(2), the special orthogonal matrices in two dimensions describing 2D-rotations.

Another subset of *O*(2) is that defined by $\det(O) = -1$. It consists of elements of *SO*(2) times the matrix

$$\begin{pmatrix} 1 & 0 \\ 0 & -1 \end{pmatrix} \tag{16}$$

corresponding to discrete transformations as

$$\begin{aligned} x &\to x \\ y &\to -y \end{aligned} \tag{17}$$

which takes a plane and maps it into its mirror image. For galaxies, the two *enantiomers* are *mirror* trailing and leading structures and, in order to pass from one to the other, a discrete inversion is necessary (Figure 2). In this sense, spiral galaxies are *non-superimposable chiral objects*.

Another important remark has to be done at this point. Let us take a complex number $u = a + ib$. It can be transformed as

$$u' = U(\vartheta)u = e^{i\vartheta}u \tag{18}$$

where $U(\vartheta)$ is a complex unitary matrix

$$U \times U^\dagger = \mathbf{I} \tag{19}$$

The set of all one-dimensional unitary matrices $U(\vartheta) = e^{i\vartheta}$ is the group $U(1)$ where the multiplication law

$$e^{i\vartheta} e^{i\vartheta'} = e^{i\vartheta + i\vartheta'} \tag{20}$$

holds. This multiplication law is the same of *O*(2) even though this construction is based on a new space of complex one-dimensional numbers. Then, we have the correspondences

$$SO(2) \approx U(1), \qquad e^{\vartheta \cdot \tau} \leftrightarrow e^{i\vartheta} \tag{21}$$

which means that two real numbers (or functions) which transform under *O*(2) can be combined into a single complex number (or function) which transforms under *U*(1).

As it is well known, *U*(1) group describes the transformations of the electromagnetic field. In the present context, it describes the symmetry between trailing and leading modes which transform each other by complex coniugation. In fact, if $U(\vartheta) = e^{i\vartheta}$ describes a trailing mode, the complex conjugate one, $U^\dagger(-\vartheta) = e^{-i\vartheta}$, represents a leading mode.

# ENVIRONMENTAL EFFECTS AND EVOLUTION OF SPIRAL GALAXIES TOWARD ACHIRAL SYSTEMS

Summarizing the above discussion, spiral galaxies (at least the grand-design ones) are enantiomers which *naturally* appear in the left-helical series characterized by trailing arms. These structures are *quasistatic density waves* which are stable with respect to the typical age of a galaxy ($\sim 10^{10}$ yr) and take part to the global motion around the bulge with periods of the order of $10^8$ yr. These results constitute the solution, obtained by Lin and Shu,[10] of the so-called *winding problem* which led to the paradox that *material arms*, composed by a fixed set of identifiable stars and gas clouds, would necessarily *wind up* on a short time scale if compared to the age of the galaxy. Considering density waves with a *global pattern speed*,[9,11] the spiral structure results stable against small perturbations and remains permanent during the evolution unless the galaxy would undergo strong gravitational interactions. This feature of a spiral galaxy can be considered *genetic* in the sense that it is the original angular momentum of the protogalaxy that gives rise to the spiral structure, which remains a permanent feature of the system. Joined to the fact that, observations almost always reveal trailing arms,[12] it seems that a *chiral asymmetry* in spiral galaxies exists and only one modality is favoured likewise in the case of aminoacids and neutrinos (both left-handed).

This statement deserves a wide discussion, first of all in connection to the other morphological types of Hubble sequence (Figure 3) and the environmetal effects which could lead toward a progressive loss of *chirality* in galaxies.

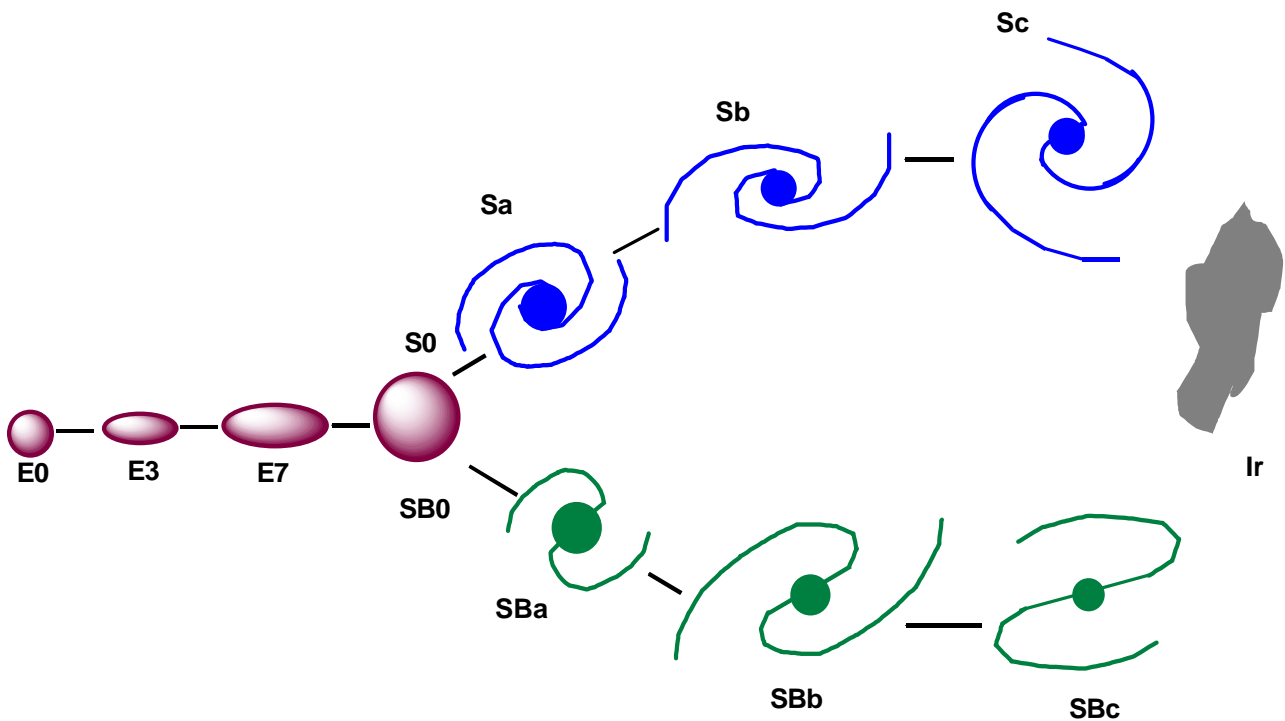

**Fig. 3.** Hubble's tuning-fork diagram of galaxy types.

Let us start our considerations pointing out a recent study by Kondepudi and Durand,[13] who performed a fine statistical analysis taking into account the spiral galaxies in the Carnegie Atlas of Galaxies.[14] They considered 540 galaxies classified as normal (**S**) or barred (**SB**) spirals. An interesting dominance of trailing structures emerges for **S**-type galaxies and a dominance of leading structures is revealed for **SB**-type galaxies. In other words, the presence of the bar is correlated to the leading modes, while grand-design normal galaxies are trailing. This fact can be interpreted considering the relative positions of morphological Hubble types in the clusters of galaxies and the probability that galaxies suffer strong or tidal gravitational interactions among them. A wide number of studies[9,15] reports that elliptical (**E**) and lenticular (**S0**) galaxies reside in large clusters, typically containing from $10^2$ to $10^3$ galaxies, while spiral galaxies are found on the boundary of clusters, in loose groups or in open fields. Clusters are often dense environments, bound together by the mutual gravitational attraction of constituent galaxies which is going to increase toward the center. Symmetric, well-shaped clusters are always core-dominated by huge elliptical **cD** galaxies (with masses ranging from $10^{13}$ to $10^{14}$ solar masses) which site in the center of symmetry (Figure 4). On the other hand, spiral galaxies reside in extremely less dense zones, where mutual gravitational interactions are very rare. In general, **E** and **S0** galaxies are classified as *cluster galaxies*, while **S** galaxies are classified as *field galaxies*.[9] The **SB** galaxies are in an intermediate situation. Bars are the result of gravitational interactions not so strong to destroy the global spiral structure. As shown by numerical simulations by Elmegreen et al.[16] and Gerin et al.,[17] the interaction of two galaxies gives rise to tidal tensions capable of twisting the buldge and inducing the formation of a bar. Unlike spiral arms, bars are not density waves since the majority of their stars always remains within the bar. This means that, although photometric observations suggest that bars and spiral arms seem joined, the former are leading structures in counter-rotation with respect to the spiral arms.[9] The cumulative effect, as in the case of M31 and M32, would be that a leading spiral structure results and the morphological type evolves as **S** $\Rightarrow$ **SB**. Without entering into details of bar formation, **SB** galaxies can be seen as an intermediate state between chirally-defined systems (**S**-galaxies) and system which are loosing, by gravitational interactions, their chiral features (**E** and **S0** galaxies) since they are living in much dense environments. Gravitational field and tidal encounters act as a sort of thermic or photo-induced racemization for enantiopure compounds.[18] In other words, starting from a well-defined chiral modality (trailing **S**-galaxies), an evolution through a *quasi-racemate*[7] (**S** and **SB**-galaxies) and toward completely achiral objects (**E**, **S0** galaxies) can be envisaged.

This process can be viewed as a *dynamical loss of chirality*. A pictorial scheme of the situation is proposed in Figure 4, where the *genetic chirality* is going to be lost when the systems are evolving toward high density zones.

As concluding remark, it appears that the Hubble morphological sequence is related to the degree of chirality of galaxies and, essentially, we can say that late-type galaxies (**S**) are chiral objects, while early-type galaxies (**E**) are achiral objects. In this perspective, the Hubble diagram can be seen as a chirality loss sequence indicating the dynamical evolution of the stellar systems.

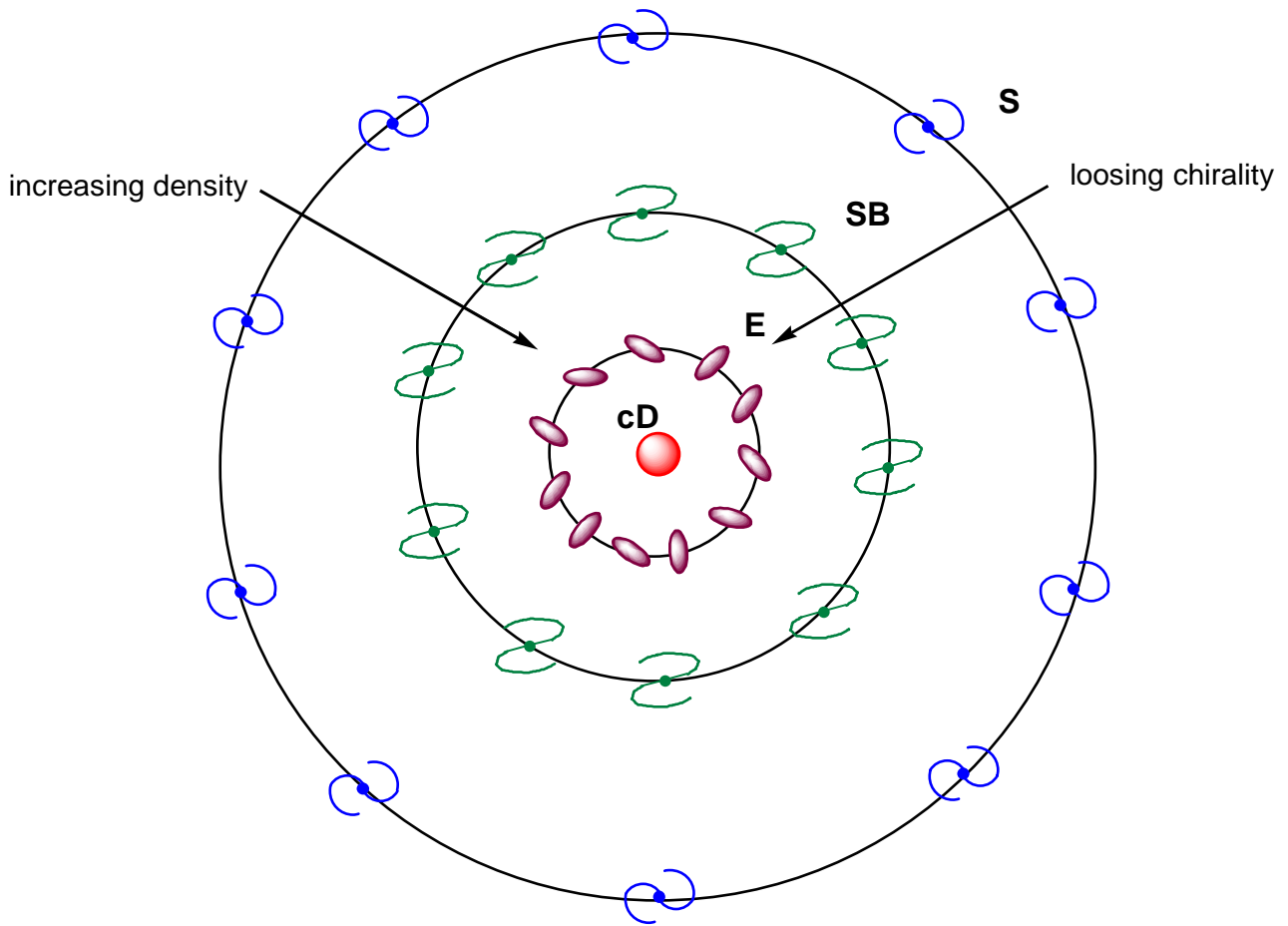

**Fig. 4** Pictorial view of morphology-density relation indicating the loss of chirality toward the centre of clusters.

# DISCUSSION AND OUTLOOK

Galaxies can be considered chiral systems and the *chirality degree* depends on environmental effects. Isolated, weak gravitationally interacting galaxies are *more* chiral than galaxies in highly dense environments as clusters. The *full* chiral systems are spiral galaxies which show 2D-chirality, algebraically described by the *O*(2) orthogonal group of rotations and inversions. Observational data indicate a preferred left-handed chirality for those normal spirals where the rotation of the arms can be unambiguosly defined (trailing modes) as the isolated grand-design spirals. Barred spirals can be the results of tidal encounters with retrograde-moving objects and they always show a chiral characterization (an excess of leading modes). This morphology-density relation indicates that the same Hubble sequence can be specified by chirality. In other words, chirality looks like a secondary feature able to denote a *segregation* between early-type and late-type stellar systems.

However, this preliminary result on a preferred genetic modality of chiral galaxies has to be more confidentially supported by deeper and refined analysis of large observational data sets (for example 2dF or SDSS surveys),[19] as soon as their morphological and dynamical characters will be completely defined. In particular, the use of deep and wide astronomical surveys (containing data between 100,000 and 1,000,000 of galaxies with rotation curves coupled with the overall angular momentum) can greatly enhance the statistics to support our result.

At this point, an important concluding remark is necessary on the underlying philosophy of our approach. Clearly, the physics of chirality in two dimensions (spiral galaxies) is quite different from that in three dimensions (tetrahedral molecules) or from that of elementary particles. The unifying view which we propose is due to the fact that all such classes of objects can be dealt with orthogonal group *O*(N), where N is the number of contituents, which determines the number of dimensions in an "abstract" space of configurations. N = 2 is the number of arms for galaxies, or the number of polarization states for particles (e.g. neutrinos), N = 4 is the number of ligands for a tetrahedron. These numbers are independent of the "true" physical space where objects lives. Furthermore, as we discussed above, N! is the number of configurations that the object can assume in such an abstract space (e.g. the number of 24 Fischer projections or the 2 trailing and leading modes of galaxies), while $\frac{N(N-1)}{2}$ is the number of independent parameters which specifies the *O*(N) group. In the case of tetrahedral molecules is six, in the case of galaxies is one (rotation angle). However, it is misleading to state that chirality of particles, molecules and galaxies is the "same" fundamental feature for so widely different systems. Our viewpoint, is that the analogy[4] of dealing with different objects under the standard of $O$(N) groups should be seriously considered in order to achieve a final and fundamental theory of chirality suitable of working at any scale and for any physical space (2D or 3D). A great deal of work in this direction has to be done.

## ACKNOWLEDGMENTS

The authors thank the referee whose comments allowed to clarify several important points of the paper.